\documentclass[journal]{IEEEtran}

\ifCLASSINFOpdf
\else
\fi
\usepackage{subfiles}
\usepackage[dvipdfmx]{graphicx}
\usepackage{color}
\usepackage{cite}
\usepackage{amsmath,amssymb,amsfonts,cases,empheq,amsthm}
\usepackage{enumitem}
\usepackage{hyperref}
\usepackage{subfigure}
\usepackage{fvextra} 
\usepackage{listings}
\usepackage{tcolorbox}

\newcommand\Colored[2]{{\color[rgb]{#1}#2}} 
\newcommand\ColorA[1]{\Colored{0,0,0}{#1}}
\newcommand\ColorB[1]{\Colored{0,0,0}{#1}}

\newcommand\LineBreak{\slash n}

\theoremstyle{remark}
\newtheorem{remark}{Remark}
\newtheorem{example}{Example}

\newcommand\RR{\mathbb{R}}

\newcommand\RouterSet{\mathcal{N}}
\newcommand\LinkSet{\mathcal{L}}
\newcommand\UserSet{\mathcal{S}}
\newcommand\UserTraffic{\phi}

\newcommand\DCSet{\mathcal{D}}

\newcommand\TrafficCapacity{\bar\phi}
\newcommand\LinkLatency{T}

\newcommand\CPU{\mathrm{CPU}}
\newcommand\LB{\mathrm{LB}}

\newcommand\CPUParameter{\theta^\CPU}
\newcommand\LBParameter{\theta^\LB}

\newcommand\CPUActual{z^\CPU}
\newcommand\LBActual{z^\LB}

\newcommand\UpdateMarker{\Delta\theta}
\newcommand\CPUUpdateMarker{\Delta\CPUParameter}
\newcommand\LBUpdateMarker{\Delta\LBParameter}

\newcommand\CPUUpdateRate{\eta_\CPU}
\newcommand\LBUpdateRate{\eta_\LB}

\newcommand\IntentClassifierForCPU{C^\CPU}
\newcommand\IntentClassifierForLB{C^\LB}

\newcommand\CPUCapacity{{v_\CPU}}

\begin{document}
\title{Chat-Driven Optimal Management for Virtual Network Services}
\author{Yuya~Miyaoka,
        Masaki~Inoue,
        Kengo~Urata,
        and~Shigeaki~Harada
\thanks{Yuya Miyaoka and Masaki Inoue are with the Department of Applied Physics and Physico-Informatics, Keio University, Kanagawa 223-8522, Japan
(e-mail: miyaoka.yuya@keio.jp; minoue@appi.keio.ac.jp).
Kengo Urata is with Network Service Laboratories, NTT, Inc., Tokyo 180-8585, Japan. Shigeaki Harada is with IOWN Product Design Center, NTT, Inc., Tokyo 180-8585, Japan.
(e-mail: kengo.urata@ntt.com; shigeaki.harada@ntt.com)

This research was conducted while the fourth author (Harada) was affiliated with the Network Service System Laboratories.}}

\markboth{Journal of \LaTeX\ Class Files,~Vol.~14, No.~8, August~2015}%
{Shell \MakeLowercase{\textit{et al.}}: Bare Demo of IEEEtran.cls for IEEE Journals}
\maketitle
\begin{abstract}
This paper proposes a chat-driven network management framework that integrates natural language processing (NLP) with optimization-based virtual network allocation, enabling intuitive and reliable reconfiguration of virtual network services. Conventional intent-based networking (IBN) methods depend on statistical language models to interpret user intent but cannot guarantee the feasibility of generated configurations. To overcome this, we develop a two-stage framework consisting of an Interpreter, which extracts intent from natural language prompts using NLP, and an Optimizer, which computes feasible virtual machine (VM) placement and routing via an integer linear programming. In particular, the Interpreter translates user chats into update directions, i.e., whether to increase, decrease, or maintain parameters such as CPU demand and latency bounds, thereby enabling iterative refinement of the network configuration.
In this paper, two intent extractors, which are a Sentence-BERT model with support vector machine (SVM) classifiers and a large language model (LLM), are introduced. Experiments in single-user and multi-user settings show that the framework dynamically updates VM placement and routing while preserving feasibility. The LLM-based extractor achieves higher accuracy with fewer labeled samples, whereas the Sentence-BERT with SVM classifiers provides significantly lower latency suitable for real-time operation. These results underscore the effectiveness of combining NLP-driven intent extraction with optimization-based allocation for safe, interpretable, and user-friendly virtual network management.
\end{abstract}


\IEEEpeerreviewmaketitle

\section{Introduction}
\IEEEPARstart{S}{erver} virtualization technology enables the construction of multiple virtual machines (VMs) on a single physical server, thereby achieving efficient utilization of computing resources and flexible server operation \cite{Daniels09_Server}. In addition, Network Function Virtualization (NFV) allows network functions such as routers, firewalls, and load balancers to be virtualized as software running on general-purpose servers \cite{Yi18_Comprehensive}. With these developments, network configurations including the setup of routing paths and the placement of VMs can be flexibly controlled, allowing for the provision of various types of virtual network services. A representative service model is Infrastructure as a Service (IaaS \cite{Whaid14_IaaS}), which provides not only computing resources such as servers and storage but also a virtual network infrastructure. This enables users to deploy their services without building or maintaining physical infrastructure, thereby enhancing their development efficiency.

The process of providing such virtual network services can be broadly divided into two stages. The first stage is to define network requirements based on user demands (intents), including resource capacity, bandwidth, and latency. The second stage is to appropriately control the placement of VMs across data centers and the transmission paths between users and VMs so as to satisfy the defined requirements. While many studies have addressed the latter stage by proposing control methods that ensure optimality and efficiency \cite{Schardong21_NFV,Huin17_Optimization}, the former stage of requirement specification still relies heavily on manual effort. In practice, this stage can account for more than half of the total service provisioning time. Therefore, the automation of the requirements specification process is a critical challenge for realizing rapid provisioning of virtual network services.

To address this challenge, Intent-Based Networking (IBN) has recently attracted increasing attention as a promising technology for end-to-end network operations, driven by user intent \cite{Manias24_Semantic,Mcnamara23_NLP,Jacobs21_Lumi,Tu25_Intent}. Specifically, it automates the process from defining network requirements based on user-intent information expressed in natural language to executing network control.
This comprehensive automation significantly reduces the time from receiving user-intent information to executing control, thereby accelerating service provisioning. Since users can obtain optimal network services simply by communicating their desired outcomes in natural language, the IBN also provides intuitive operability and high reliability, ultimately enhancing user satisfaction. From the perspectives of automation, accelerated service provisioning, and assurance of overall system reliability, we need to incorporate user-intent information into network operations.

In conventional IBN, statistical approaches have been introduced to flexibly interpret diverse user intents, thereby expanding the possibilities of virtual machine services. However, purely statistical methods may fail to ensure the feasibility of the resulting network configurations, raising concerns about safety and reliability. To guarantee feasibility, model-based (e.g., optimization-based) methods, such as those proposed in \cite{Schardong21_NFV,Huin17_Optimization,Urata24_Distributionally,Alameddine17_Scheduling,Carpio17_VNF}, remain desirable.
This paper aims to integrate the strengths of both paradigms—statistical and model-based approaches. Building on natural language processing (NLP) techniques, we develop a novel framework that not only accommodates diverse user instructions but also guarantees the feasibility of the resulting network configurations.

\begin{figure}[t]
    \centering
    \includegraphics[width=1\linewidth]{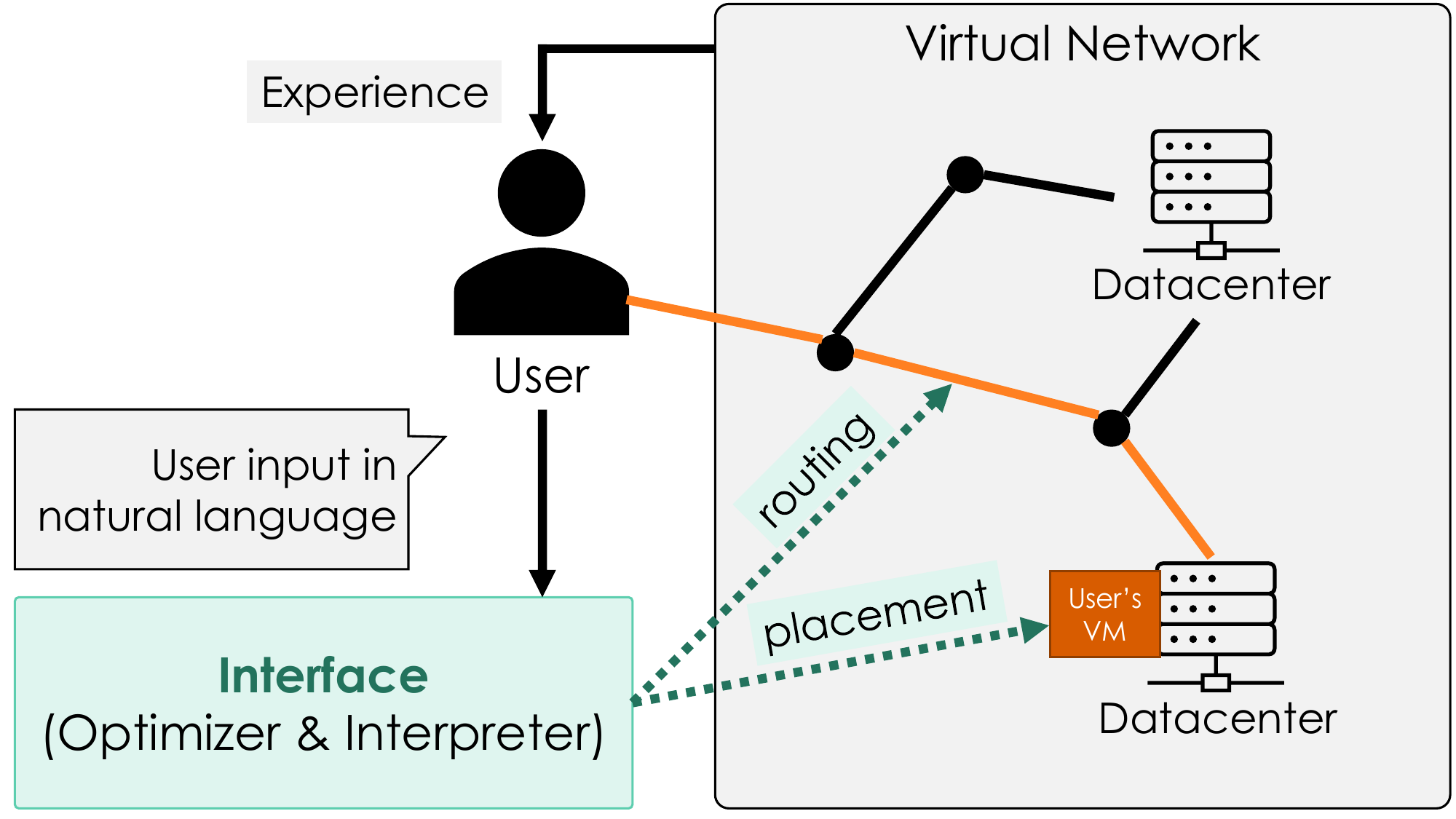}
    \caption{Conceptual structure of proposed framework}
    \label{F:I.Concept}
\end{figure}

\begin{figure}[t]
    \centering
    \includegraphics[width=1\linewidth]{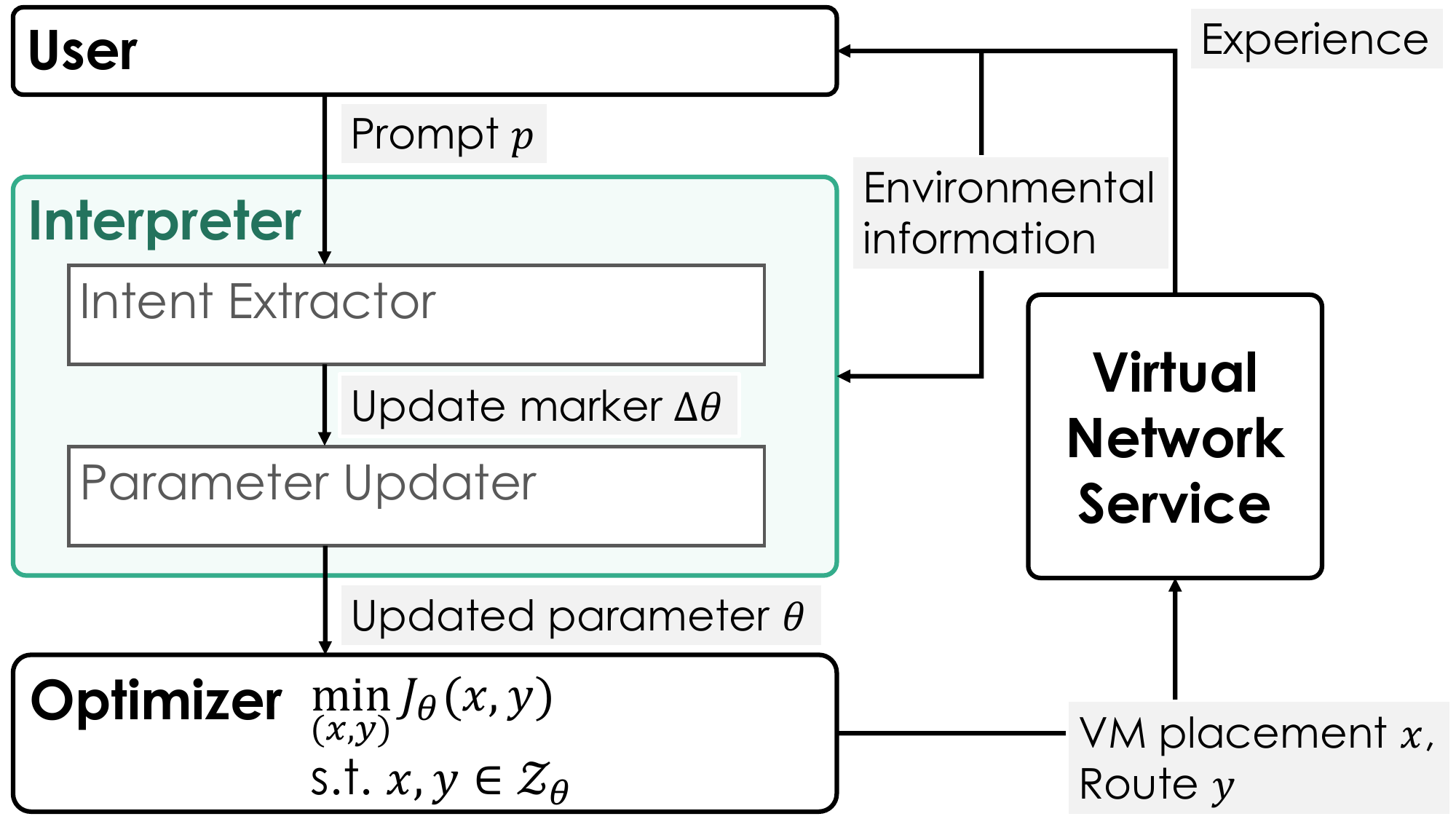}
    \caption{Detailed structure of the overall system}
    \label{F:I.Structure}
\end{figure}

This paper presents a framework that leverages NLP to enhance the accessibility and intuitiveness of virtual network services. Building on our previous research \cite{Miyaoka24_Chatmpc}, we design an interface that bridges user interactions and the virtual network allocation process, as illustrated in Fig.~\ref{F:I.Concept}. Through this interface, the system receives user instructions expressed in natural language and incorporates them into VM placement and routing decisions.
As shown in Fig.~\ref{F:I.Structure}, the interface consists of two key components: an Interpreter and an Optimizer. The Interpreter analyzes user chat using NLP models and converts linguistic expressions into structured specification parameters such as CPU demand and latency requirements that drive the network allocation problem. To accommodate diverse user expressions, we develop two Interpreter architectures: one combining Sentence-BERT models with Support Vector Machine (SVM) classifiers and another leveraging a large language model (LLM). These implementations differ in computational cost and interpretive flexibility, enabling a trade-off between accuracy and operational efficiency.
The Optimizer then solves an integer linear programming (ILP) formulation of the virtual network allocation problem using the Interpreter-derived specifications. By computing feasible VM placement and routing that satisfy resource, latency, and capacity constraints, the Optimizer ensures that user intent is reflected safely and optimally in the resulting network configuration.
This paper also validates the effectiveness of the proposed framework through two experiments. 
The first experiment demonstrates the system's ability to iteratively adjust VM allocation and routing in response to natural language prompts, showing how the parameter update mechanism influences the final network configuration.  
The second experiment provides a performance comparison of different intent extractor implementations in the Interpreter.

This paper is an extended version of our previous conference paper \cite{Miyaoka25_ChatDriven}. The differences from the conference version are summarized as follows:
\begin{itemize}
    \item LLM-based Interpreter: In addition to the Sentence-BERT model with an SVM-based classifier used in the conference version, we implement an LLM-based Interpreter, which offers higher expressiveness and better adaptability with limited labeled data.
    \item Arbitration for the multi-user case: We introduce an arbitration mechanism that accepts the largest possible number of user requests when multiple chat prompts arrive simultaneously and may conflict with one another. 
    \item Quantitative evaluation of the Interpreter: We quantitatively compare the Sentence-BERT + SVM Interpreter and the LLM-based Interpreter in terms of intent extraction accuracy, the amount of training data required, and computational time.
\end{itemize}

The rest of the paper is organized as follows. A comparison to related works
is presented in Section \ref{Section:Related_Works}. In Section \ref{Section:M}, we propose a novel end-to-end VM allocation framework driven by user intent expressed in natural language form, which is constructed from the Interpreter and the Optimizer. In Section \ref{SubSection:M.最適化器}, Optimizer is formulated as an ILP and in Section~\ref{SubSection:M.インタプリタ}, two types of Interpreter are introduced based on a Sentence-BERT model with SVM classifiers and LLM models. In Section~\ref{Section:E}, experiments are conducted. Finally, the paper is concluded in Section~\ref{Section:C}.

Notation: For $a \in \mathbb{R}^n$, $||a||_1$ denotes the L1 norm of $a$.

\section{Related Works}\label{Section:Related_Works}
Several prior studies have formulated the virtual network allocation problem as a numerical optimization problem to automatically determine the optimal placement of VMs \cite{Schardong21_NFV,Huin17_Optimization,Urata24_Distributionally,Alameddine17_Scheduling,Carpio17_VNF,Hosseini19_Probabilistic}. For instance, the study in \cite{Huin17_Optimization} formulates the VM placement problem as an ILP problem, whereas the studies in \cite{Urata24_Distributionally}  and \cite{Alameddine17_Scheduling} formulate it as a mixed-integer linear programming (MILP) problem. Moreover, certain works incorporate uncertainties that arise in practical environments into their formulations. Specifically, \cite{Urata24_Distributionally} considers the network system powered by renewable energy sources and proposes a robust virtual network allocation method that handles uncertain power availability. Furthermore, the paper \cite{Hosseini19_Probabilistic} assumes a Gaussian distribution of traffic demands and presents a method that constrains the probability of congestion below a given threshold.

Although these studies address the virtual network allocation problem, they operate under the implicit assumption that specification parameters such as CPU resources and latency bounds are provided. In practice, however, specifying these parameters can be difficult for users, particularly those without professional expertise. Furthermore, updating such specifications in real time during service operation remains challenging, even for users with prior experience in virtual network services.

Several works have investigated the utilization of NLP for network operation and management aiming for enhanced accessibility and intuitiveness for users \cite{Yao24_Velo, Su24_Leveraging, Houssel24_Towards, Rezaei25_Fedllmguard, Van24_Towards, Abbas24_leveraging}.
The paper \cite{Yao24_Velo} proposes a cloud–edge collaborative approach to leveraging LLMs and designs a multi-agent reinforcement learning–based decision mechanism that dynamically selects whether to query a cloud-hosted LLM or return an edge-cached LLM response. This mechanism reduces both latency and computational cost in LLM operations.
The paper \cite{Su24_Leveraging} proposes an LLM-based method for forecasting the demand of virtual network function resources, such as CPU, memory, and storage, and demonstrates higher accuracy than probability-based models.
The papers \cite{Rezaei25_Fedllmguard,Houssel24_Towards} leverage LLMs for fault and anomaly detection. In particular, \cite{Houssel24_Towards} employs an LLM to generate natural-language explanations of detection outcomes with explicit evidential support, thereby enhancing reliability and explainability of the detection system.
The papers \cite{Van24_Towards, Abbas24_leveraging} employ the in-context learning of LLMs for operating network systems. In particular, the study \cite{Abbas24_leveraging} utilizes LLMs to perform wireless symbol demodulation in limited-data scenarios. It also demonstrates that calibrated prompting, which refines LLM inputs, enables LLMs to outperform traditional DNNs. 
Across these studies, NLP/LLM components are \textit{directly} utilized for enhancing network operation. 
This paper proposes an approach that employs NLP/LLMs to extract user intents into the specification of the virtual network allocation. \ColorB{This \textit{indirect} approach} contributes to enhancing the feasibility and explainability of the resulting network configuration.

Some conventional works \cite{Lotfi25_LLM,Manias24_Semantic,Mcnamara23_NLP,Jacobs21_Lumi,Tu25_Intent,Asif25_RIBN,Habib25_LLM} employ an indirect approach that leverages NLP/LLMs for translating user intents expressed in natural language into network control specifications.. For instance, the paper \cite{Lotfi25_LLM} does not use LLM outputs to excuse the slicing control for open radio access networks. Instead, it uses LLMs to argue the state representations of the multi-agent deep reinforcement learning with semantic features. The papers leverage LLMs to \ColorA{extract} user intents expressed in natural language into network control specifications \cite{Manias24_Semantic,Mcnamara23_NLP,Jacobs21_Lumi,Tu25_Intent}. For instance, the paper \cite{Manias24_Semantic} introduces semantic routing as a pre-processing step to enhance the translation of user intent to reduce hallucinations as compared with standard LLMs. In addition, the paper \cite{Mcnamara23_NLP} proposes an intent-translation system comprising two components: an Intent Engine and an AI Engine. The latter complements the original intent with missing low-level context. Two prior studies \cite{Asif25_RIBN, Habib25_LLM} propose IBN with a three-stage strategy consisting of intent translation, intent validation, and control execution, and this strategy is conceptually similar to this paper. Here, intent validation indicates a mechanism that determines whether the \ColorA{extracted} intent is applicable under current network conditions. In both studies, LLMs are utilized for intent translation and reinforcement learning is employed for control execution; the first study \cite{Asif25_RIBN} employs a validation method based on k-nearest neighbor algorithm, whereas the second study \cite{Habib25_LLM} proposes a Transformer-based approach. As compared with conventional indirect approaches, the key contributions of this paper are outlined as follows:
\begin{itemize}
    \item By employing ILP as the control-execution module (referred to as the Optimizer in this paper), we compute optimal VM placement and routing under the specified constraints while faithfully reflecting user intent. In contrast, prior studies rely on reinforcement learning, which provides no guarantees of strict optimality.
    \item In this paper, we solve an optimization problem under the network constraints, which are parametrized by specification values obtained by translating user intents via an Interpreter. Under the proposed framework, the feasibility of the control actions required to satisfy user intents can be rigorously guaranteed through an optimization-theoretic approach. Since feasibility is guaranteed by the Optimizer, the proposed framework provides inherent robustness to LLM hallucinations even without an intent-validation module, as employed in conventional works \cite{Asif25_RIBN, Habib25_LLM}.
\end{itemize}

\section{Methodology}\label{Section:M}

\begin{table*}
    \centering
    \caption{Symbols}
    \begin{tabular}{p{0.15\linewidth}p{0.75\linewidth}}
        \hline
        Symbol & Meaning \\
        \hline
         $k$ & The chat step and denoted by $k=0,1,\ldots$.\\
         $N, \RouterSet$ & Number of routers, and set of the router index, i.e., $\RouterSet=\{1,\ldots,N\}$. \\
         $L, \LinkSet$ & Number of links, and set of the link index, i.e., $\LinkSet=\{1,\ldots,L\}.$ \\
         $D$ & Number of datacenters (DCs). \\
         $S, \UserSet$ & Number of users, and set of the user index, i.e., $\UserSet=\{1,\ldots,S\}$. \\
         $\TrafficCapacity_l$ & Bandwidth, a maximum traffic amount, of Link $l\in\LinkSet$. \\
         $\UserTraffic_s$ & Traffic volume generated by User $s$. \\
         $\LinkLatency_l$ & Latency, the time delay that occurs when transmitting the data through Link $l\in\LinkSet$.\\
         $\theta\in\RR^2$&  \\
         $\CPUParameter_s(k)\in\RR$& \textbf{Updateable parameter.} The minimum CPU resource of the VM that User $s\in\UserSet$ desires to use.\\
         $\LBParameter_s(k)\in\RR$& \textbf{Updatable parameter.} The latency bound, the maximum latency time for communication from User $s\in\UserSet$ to its VM.\\
         $\CPUUpdateRate, \LBUpdateRate$ &  Hyperparameters of the Interpreter. Determine the magnitude of the update range. The ranges are $\CPUUpdateRate>1$ and $\LBUpdateRate>1$.\\
         $\UpdateMarker\in\{-1,0,+1\}^2$ & Called ``update marker''. Values $\UpdateMarker=[\CPUUpdateMarker ~ \LBUpdateMarker]^\top$ indicate direction of update from the current parameter values, $\CPUParameter$ and $\LBParameter$, respectively.\\
         $x(k) \in \{0,1\}^{S\times D}$& \textbf{Decision variable.} Represents the VM placement. The element $x_{sd}$ takes 1 if the VM of User $s$ is allocated into the DC $d$ and otherwise 0. \\
         $y(k) \in \{0,1\}^{S\times L}$& \textbf{Decision variable.} Represents the routing. The element $y_{sl}$ takes 1 if the communication between User $s$ and the VM is passed through Link $l$ and otherwise 0.\\
         \hline
    \end{tabular}
    \label{T:M.Symbols}
\end{table*}

This section formulates the overall system design, including the Optimizer and the Interpreter. 
The Optimizer formulates the virtual network allocation problem as an ILP, which determines VM placement and routing. The objective function and the constraints of the optimization problem are parametrized by $\theta$, which represents the updateable specification of the virtual network allocation problem. 
The Interpreter estimates the user's requirements based on their chat text in natural language format, referred to as the ``prompt''.  The examples of Interpreter implementations are described in Subsection~\ref{Subsection:M.IntentExtractorImplementations}. \ColorA{The iterative process of parameter update is performed to align network services with user preferences and requirements. The chat steps are denoted by $k=0,1,\ldots$. 
The updatable parameter $\theta(k)$ and decision variables $(x(k),y(k))$ are dependent on the chat step $k$, which are often denoted as $(x,y)$.} Symbols used in this paper are shown in Table~\ref{T:M.Symbols}.

\subsection{Optimizer}\label{SubSection:M.最適化器}
We establish a simplified setting of the virtual network service and the virtual network allocation problem. In the problem, each user rents a single VM. The network topology is a directed graph where routers and links are represented as nodes and edges, respectively. \ColorA{Datacenters (DCs) are a special type of router that places VMs and have substantial compute resources like CPUs and memory.} Each user needs to rent a VM within a DC and to communicate with the VM through the links.

The decision variables of the virtual network allocation problem are $x$ and $y$, representing VM placement and routing, respectively. We define three objective functions to be minimized:
\begin{itemize}[leftmargin=1em]
    \item $J_1(x,y) = \max_{l\in\LinkSet} (y_{sl}\UserTraffic_s) / \TrafficCapacity_l$: This function represents the maximum link usage in the network. Minimizing $J_1$ aims to distribute the traffic load of each link.
    \item $J_2(x,y)=\sum_{s\in\UserSet}\sum_{l\in\LinkSet} y_{sl}\LinkLatency_l$: This function represents the total network latency. Minimizing $J_2$ prevents the occurrence of undesirable routes, such as unnecessarily long routes.
    \item \ColorA{ $J_3(x,y) = ||x - x^- ||_1 $: This function measures the change in VM placement between chat steps. The constant $x^-$ is the VM placement solution $x$ from the previous chat step, and $x^-$ at $k=0$ is given, such as $x^-=0$. Minimizing $J_3$ discourages significant and sudden changes to VM placements, thereby maintaining the stability and usability of the virtual network service during the chat steps.}
\end{itemize}

The virtual network allocation problem is formulated as the following ILP: 
\begin{subequations}
\begin{align} 
    \min_{x,y} ~& \omega_1 J_1(x,y) + \omega_2 J_2(x,y) + \omega_3 J_3(x,y),\\
    \text{s.t.}~
    & x\in\{0,1\}^{S\times D} , y\in\{0,1\}^{S\times L}, \label{E:M.VNAP.Binary} \\ 
    & \sum_{s\in\UserSet} y_{sl}\UserTraffic_s \le \TrafficCapacity_l, \forall l\in\LinkSet, \label{E:M.VNAP.Congestion} \\ 
    & \sum_{s\in\UserSet} x_{sd}\CPUParameter_s \le \CPUCapacity_d, \forall d\in\DCSet, \label{E:M.VNAP.CPUReq}\\ 
    & \sum_{l\in\LinkSet} y_{sl}\LinkLatency_l \le \LBParameter_s, \forall s\in\UserSet, \label{E:M.VNAP.LBReq} \\ 
    & \sum_{d\in\DCSet} x_{sd}=1, \forall s\in\UserSet, \label{E:M.VNAP.Singleton} \\ 
    & \Pi_n^\top y_s =  
        \begin{cases}
            +1,& n=n_s, n \text{ is not DC}, \\
            -1 ,& n\ne n_s, n\text{ is DC}, \\
            0, & \text{else},
        \end{cases} \label{E:M.VNAP.Kirchhoff}\\
    & \forall n\in\RouterSet, \forall s\in\UserSet.  \notag
\end{align}
\label{E:M.VNAP}
\end{subequations}

Symbols $\omega_1$, $\omega_2$, and $\omega_3$ are weighting constants for the objective functions.
Constraint \eqref{E:M.VNAP.Congestion} ensures to prevent the congestion on each link $l\in\LinkSet$. Constraint \eqref{E:M.VNAP.CPUReq} ensures to limit the total CPU resources on each DC $d\in\DCSet$, satisfying the user requirements on CPU resources, $\CPUParameter$. Constraint \eqref{E:M.VNAP.LBReq} ensures to satisfy user requirements on the latency bound $\LBParameter$. Constraint \eqref{E:M.VNAP.Singleton} ensures that each user owns only one instance of VM. 
\ColorA{In constraint \eqref{E:M.VNAP.Kirchhoff}, $\Pi_n$ is the incidence matrix of the network topology at $n$-th row, and $y_s$ is the decision variable at $s$-th row, indicating the link usage of the user $s$. Constraint \eqref{E:M.VNAP.Kirchhoff} is for creating a one-stroke route from each user to the DC that allocates the user's VM, ensuring the incoming and outgoing data balance for each router.}

\begin{remark}
    Problem \eqref{E:M.VNAP} is formulated in a simplified setting and can be further extended. Extensions include considering storage capacity constraints or integrating the power consumption for data transmission accross links into the objective function.
\end{remark}

Note that Problem \eqref{E:M.VNAP} involves the parameter, $\CPUParameter_s$ and $\LBParameter_s$ for each users $s\in\UserSet$. The parameter of User $s$ is denoted as $\theta_s:=[\CPUParameter_s ~ \LBParameter_s]^\top$.

\subsection{Interpreter}\label{SubSection:M.インタプリタ}
An Interpreter is the key component of the presented service system, which is illustrated in Fig.~\ref{F:I.Structure}. The architecture of an Interpreter consists of a cascaded structure with two units: an intent extractor and a parameter updater.

\ColorA{
In an intent extractor, the user-provided prompt $p$ is analyzed to output the update marker $\UpdateMarker=[\CPUUpdateMarker ~ \LBUpdateMarker]^\top\in\{-1,0,+1\}^2$. The values $\CPUUpdateMarker\in\{-1,0,+1\}$ and $\LBUpdateMarker\in\{-1,0,+1\}$ indicate the direction of update from the current parameter value: $-1$ indicates to reduce the value, $+1$ indicates to increase the value, and $0$ indicates to maintain the value. The symbol $\UpdateMarker$ is called ``update marker''.  An intent extractor is responsible for analyzing the prompt to determine whether each topic should be increased, decreased, or maintained. Specifically, it outputs $+1$ / $-1$ if the prompt contains a request to increase/decrease the value, $0$ if the topic is irrelevant or the prompt is uninterpretable. Two implementations of the intent extractor are presented in Subsection~\ref{Subsection:M.IntentExtractorImplementations}.

In parameter update, the parameter $\theta(k)$ is updated based on the determined update marker $\UpdateMarker$, as follows:
} 
\begin{subequations}
\begin{empheq}[left={\empheqlbrace}]{alignat=1}
    \CPUParameter_s(k+1) &= \CPUUpdateRate^{\CPUUpdateMarker(k)} \CPUActual\\
    \LBParameter_s(k+1) &= \begin{cases}
        \LBUpdateRate \LBParameter_s(k) ,& \LBUpdateMarker(k)=+1,\\
        \LBParameter_s(k) ,& \LBUpdateMarker(k)=0,\\
        \LBUpdateRate^{-1} \LBActual_s(k) ,& \LBUpdateMarker(k)=-1,
    \end{cases}
\end{empheq}\label{E:M.ParameterUpdate}\end{subequations}
The actual values $\CPUActual$ and $\LBActual$ must be measured from a real-world network system.

\begin{remark}
    In this problem addressed in this section, it is assumed that two specification parameters related to CPU resource and latency bound are configurable. The problem can be extended to include other specifications as parameters, such as storage volume, memory volume, and maximum power consumption. The additional parameters are also updated in a manner similar to the parameter update rule \eqref{E:M.ParameterUpdate} with appropriate markers.
\end{remark}

\ColorA{
\begin{remark}
When the Interpreter indicates no change to the latency bound, i.e., $\LBUpdateMarker=0$, we maintain the previous parameter value: $\LBParameter_s(k+1)=\LBParameter_s(k)$. This means that the system only updates the latency-bound parameter when the user explicitly prompts it. This ``non-interfering strategy'' is applied to prevent the service from inadvertently narrowing the feasible solution space for the virtual network allocation problem without explicit user intent. 
\end{remark}
} 

\subsection{Intent Extractor Implementations}\label{Subsection:M.IntentExtractorImplementations}
\ColorA{
This subsection shows two distinct architectures for implementing an intent extractor. The first architecture combines a high-performance natural language model with classical machine learning classifiers. The second architecture leverages the capabilities of a modern LLM.

The first architecture consists of a BERT model\footnote{A Sentence-BERT model \textit{sentence-transformers/all-MiniLM-L6-v2} on HuggingFace \cite{Reimers19_SentenceBERT} is applied in this paper.} and SVM models (BERT+SVM). The BERT model is one of the natural language processing foundation models that takes a text to output the embedding vector $e_s\in\RR^m$. The embedding vector $e_s$ is in high-dimensional space, where $m\approx1000$, and it expresses the intent of the text through its direction \cite{Reimers19_SentenceBERT}. The embedding vector $e_s$ is then fed into the following two Support Vector Machine (SVM) classifiers to determine the update marker $\UpdateMarker$ output.
\begin{itemize}[leftmargin=1em]
    \item The intent classifier for CPU resource $\IntentClassifierForCPU:\RR^m \to \{-1,0,+1\}$ classifies the embedding vector $e_s$ to select an operation for the CPU resource requirement parameter $\CPUParameter$. The classifier outputs $\IntentClassifierForCPU(e_s)=-1$ represents the decrease of the CPU resource, $\IntentClassifierForCPU(e_s)=+1$ represents the increase, and  $\IntentClassifierForCPU(e_s)=0$ represents that the prompt has no claims about CPU.
    \item The intent classifier for latency bound $\IntentClassifierForLB:\RR^m\to\{-1,0,+1\}$ classifies the embedding vector $e_s$ to select and operation for latency bound requirement parameter $\LBParameter$. The classifier outputs $\IntentClassifierForLB(e_s)=-1$ represents the decrease of the latency bound, $\IntentClassifierForLB(e_s)=+1$ represents the increase, and $\IntentClassifierForLB(e_s)=0$ represents that the prompt has no claims about the latency bound.
\end{itemize}
Both SVM classifiers are trained on pre-labeled training samples.
The final output, the update maker $\UpdateMarker$, is obtained by combining the classification results: $\UpdateMarker = [\IntentClassifierForCPU(e_s) ~ \IntentClassifierForLB(e_s)]$. 
The BERT+SVM architecture is expected to require fewer computational resources and to produce formed outputs.

The second architecture employs an LLM to perform the intent analysis and to generate the update marker $\UpdateMarker$. In this paper, the Meta Llama~3-8b-Instruct model \cite{Dubey24_Llama3} is applied. The process is as follows: 
The provided prompt is inserted into a template text that can also insert several pre-labeled training samples for few-shot learning. The pre-labeled training samples are detailed in Section~\ref{Section:E}. The combined text is then fed into the LLM model to generate a response text. Since the LLM is instructed to format the update marker information in JSON, we then perform JSON parsing to make the update marker $\UpdateMarker$.
The LLM architecture is expected to require less training data to adapt to the domain.
The template text of the LLM input is shown in Appendix~\ref{Appendix:LLMInputTemplate}.
} 

The performance comparison of two implementations is demonstrated in Section~\ref{Section:E}.

\section{Experiments}\label{Section:E}
This section outlines the entire flow of our proposed network, which involves sending prompts, solving the virtual network allocation problem, and subsequently serving virtual network services to users. 
The first two subsections outline the whole process, from processing user prompts to serving the virtual network services. The final subsection provides a performance comparison of different intent extractor implementations in the Interpreter. 

\subsection{Network Operation : Single-User Case}\label{Subsection:E.SingleUserCase} 
The first example demonstrates the network operation in a single-user case.

\begin{figure}
    \centering
    \includegraphics[width=1\linewidth]{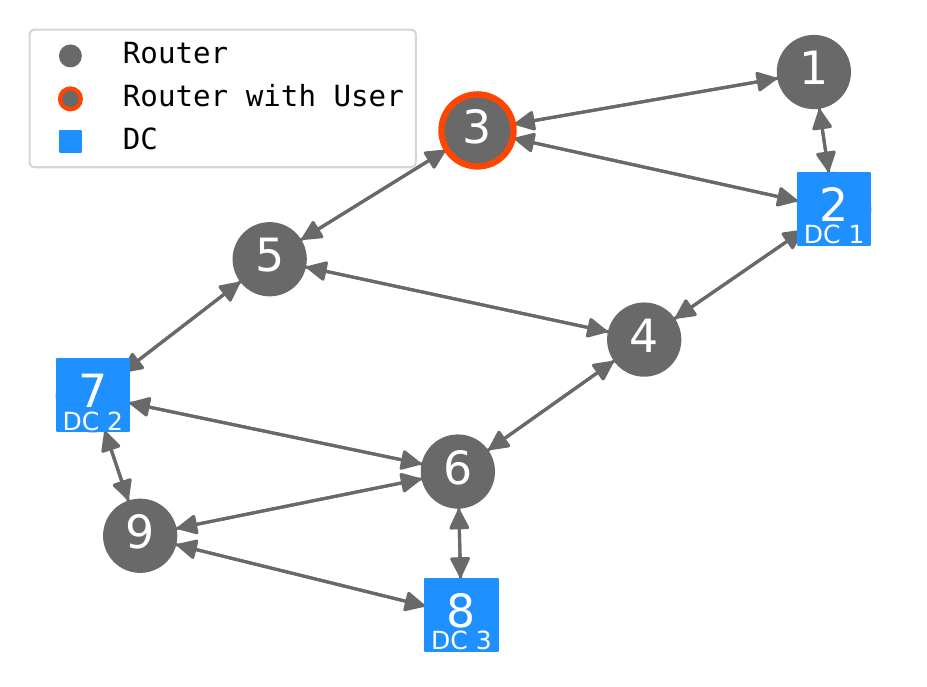}
    \caption{Network setting in the experiments}
    \label{F:E.A.Setting}
\end{figure}

We assume a single user who rents a VM from the presented virtual network, as shown in Fig.~\ref{F:E.A.Setting}. The topology used in this experiment is based on the IP network in Internet2 \cite{Summerhill06}. The network has $L=26$ links. The bandwidth of each link, $\TrafficCapacity_l$, is set to $1.0~\mathrm{Gbps}, l\in\LinkSet$ and the latency of each link, $\LinkLatency_l$, is set to $1.0~\mathrm{ms}$. The user is located at Router 3, and traffic volume, $\UserTraffic$, is $0.5~\textrm{Gbps}$.
The service provider offers solutions $(x,y)$ that update the user's network service in response to the prompts. The initial parameters are set to a CPU resource of $\CPUParameter(0)=2~\mathrm{core}$ and a latency bound of $\LBParameter(0)=3~\mathrm{ms}$. The hyperparameters of the Interpreter, as defined in \eqref{E:M.ParameterUpdate}, are set to $[\CPUUpdateRate, \LBUpdateRate]=[2, 1.5].$ The weights in the virtual network allocation problem \eqref{E:M.VNAP} are set to $\omega_1=1$, $\omega_2=0.01$, and $\omega_3=0.05$.
\ColorA{
We applied the LLM in the intent extractor of the Interpreter, as presented in \ref{Subsection:M.IntentExtractorImplementations}. To perform the few-shot learning, 30 pre-labeled examples are included in the LLM input text. The detail is shown in Subsection~\ref{Subsection:E.IntentExtractionComparison}.
}

The user provided prompts in the following order:
\begin{itemize}
    \item $k=0$ ``I no longer need such as plenty CPU''
    \item $k=1$ ``I would like to have lower latency network''
    \item $k=2$ ``Get more CPUs, please.''
\end{itemize}

\begin{figure*}[t]
  \begin{minipage}[b]{0.3\linewidth}
    \centering
    \subfigure[VM placement and routing at time $k=0$]{
        \includegraphics[width=\columnwidth]{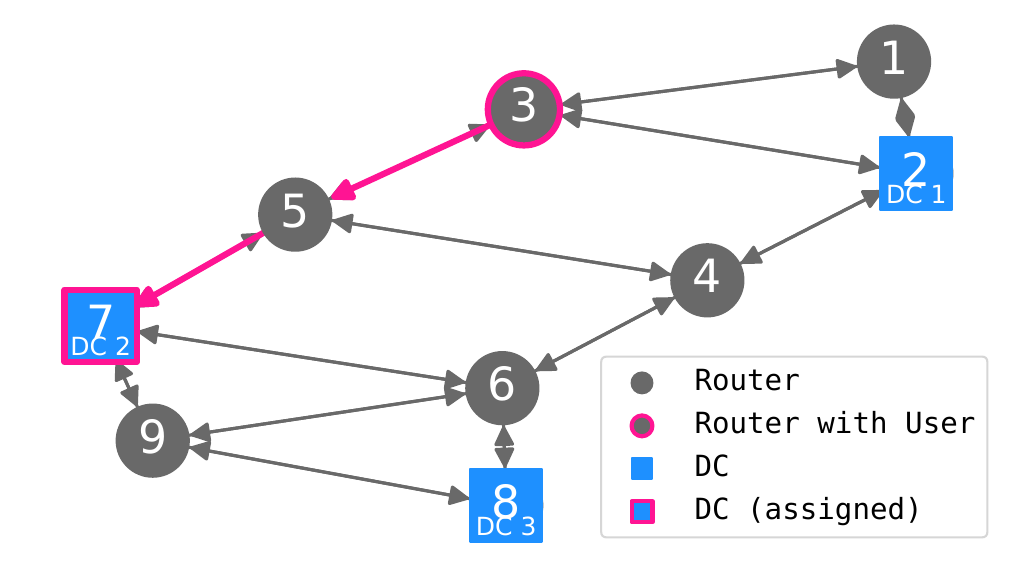}
        \label{F:E.A.Route.Time0}
    }
  \end{minipage}
  \begin{minipage}[b]{0.3\linewidth}
    \centering
    \subfigure[VM placement and routing at time $k=1$]{
        \includegraphics[width=\columnwidth]{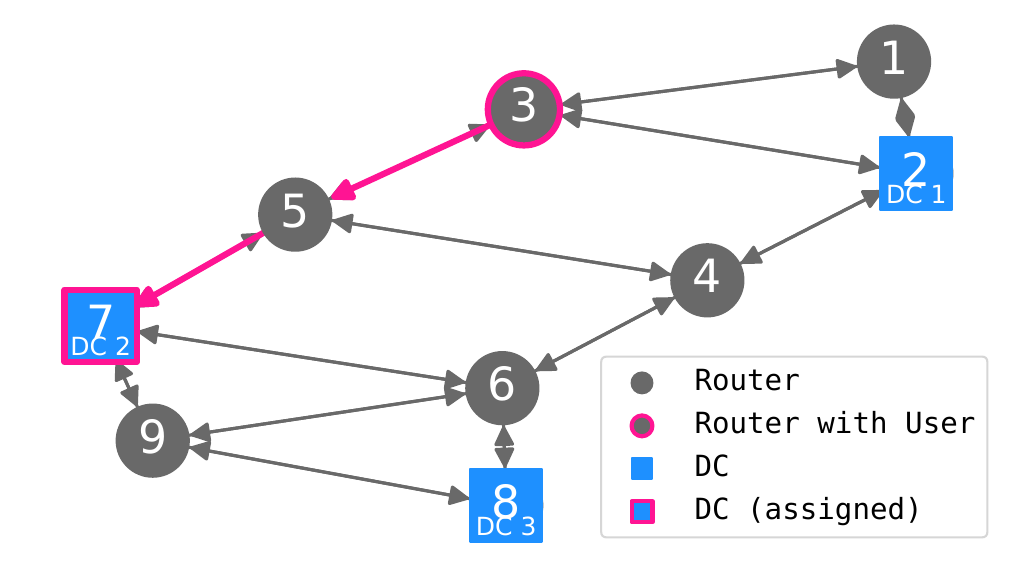}
        \label{F:E.A.Route.Time1}
    }
  \end{minipage}
  \begin{minipage}[b]{0.3\linewidth}
    \centering
    \subfigure[VM placement and routing at time $k=2$]{
        \includegraphics[width=\columnwidth]{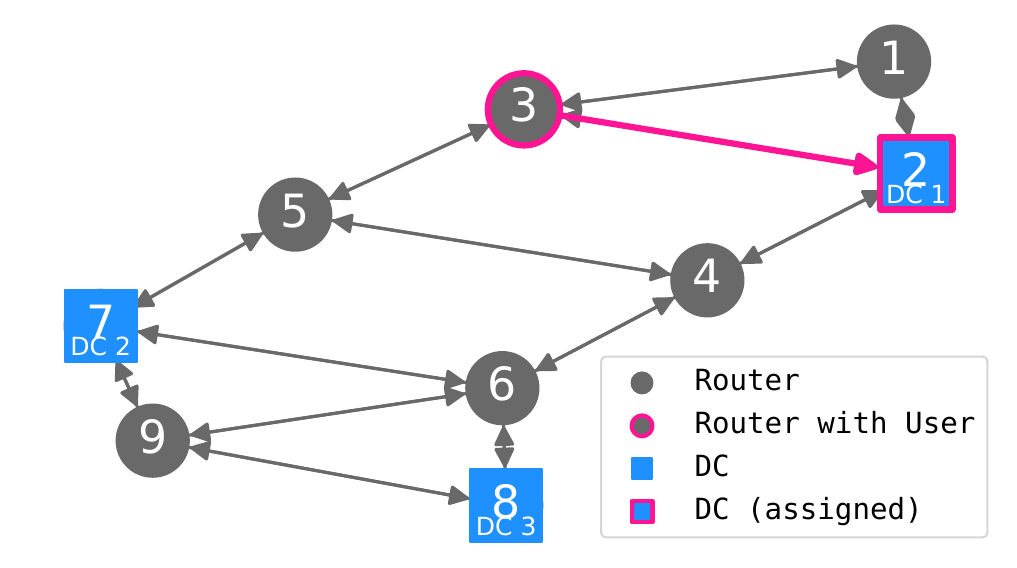}
        \label{F:E.A.Route.Time2}
    }
  \end{minipage}
  
  \caption{VM placement and routing in the single-user case}
  \label{F:E.A.Results}
\end{figure*}

The solution $y$, representing the routing, is illustrated in Fig.~\ref{F:E.A.Results}. This figure displays the DC assignment and routing. Note that the actual CPU resource assigned is equivalent to the CPU resource parameter, i.e., $\CPUActual=\CPUParameter$. The latency-bound parameter $\LBParameter$ and resulting actual latency time $\LBActual$ at each chat step $k$ is shown in Table~\ref{T:E.A.Results}.

\begin{table*}[t]
    \centering
    \caption{Results in the single-user case}
    \begin{tabular}{cccc}
    \hline
         Chat step $k$& Parameter $\LBParameter(k)~\mathrm{[ms]}$  & Actual latency $\LBActual~\mathrm{[ms]}$ & Actual CPU usage $\CPUActual(k)~\mathrm{[core]}$ \\
    \hline
         $0$ & $3.0$  & $2.0$ & $2.0$\\
         $1$ &  $3.0$ & $2.0$ & $1.0$\\
         $2$ &  $1.3$ & $1.0$ & $1.0$\\
         $3$ & $1.3$ & (Infeasible) & $2.0$\\
    \hline
    \end{tabular}
    \label{T:E.A.Results}
\end{table*}

At the initial step $k=0$, the VM is placed in DC 2, connected to Router 7.
At step $k=1$, the CPU usage declined, as intended by the prompt.
At step $k=2$, the VM is moved to DC 1, connected to Router 2, as expected. Since the prompt is related to a latency-bound topic, the latency-bound parameter was set to $\LBParameter(2)=1.3~\mathrm{ms}$, which is derived from the parameter update rule \eqref{E:M.ParameterUpdate}, specifically $ \LBParameter(2) = \LBUpdateRate^{-1} \LBActual(1)=1.5^{-1} \times 2.0~\mathrm{ms}$.
At the final step $k=3$, the optimization problem is infeasible, meaning that the virtual network problem under the parameter $\theta(3)$ is not executable and the system cannot align with the user request.
\ColorB{Unlike an end-to-end approach, our proposed framework, through the cascaded connection of the Interpreter and the Optimizer, successfully detects the infeasibility beforehand, which contributes to the preservation of network integrity.}

\subsection{Network Operation : Multi-User Case}\label{Subsection:E.MultiUserCase} 
This subsection extends the single-user case, presented in Subsection~\ref{Subsection:E.SingleUserCase}, to a multi-user case, where three users interact with the system. Users 1, 2, and 3 are located in Router 1, 3, and 4, respectively. The network's link configuration, DCs, and topology remain the same as in the single-user case. 

\ColorB{
In this scenario, multiple users may send their prompts \textit{at the same chat step}. 
This setup models cases where the order of prompts is ambiguous, i.e., multiple users send prompts almost simultaneously, or the Optimizer's execution time is long, leading to a backlog of multiple prompts accumulated before the next optimization execution.
In such cases, the parameter update rule \eqref{E:M.ParameterUpdate} is iteratively applied for each accumulated prompt within a single chat step.
Furthermore, when reflecting all accumulated prompts leads to an infeasible solution, the Interpreter in this experiment performs \textit{arbitration} to accept the largest possible number of user prompts.
Let $M$ prompts be received at the chat step $k$, and further let $\UpdateMarker_m$ be the update marker corresponding to the received prompt $p_m, m~\in\{1,\ldots,M\}$. We introduce a binary decision vector $a\in\{0,1\}^M$, where $a_m=1$ indicates acceptance of the $m$-th prompt and $a_m=0$ indicates rejection. Then, the arbitration is expressed as:

\begin{subequations}
\begin{align}
    \max_{a\in\{0,1\}^M} ~& || a ||_1  \\
    \begin{split}
            \text{s.t.} ~& \text{The virtual network allocation problem } \\ 
            & \text{with \eqref{E:M.ParameterUpdate} and} ~ \UpdateMarker = [\UpdateMarker_1 ~ \cdots ~ \UpdateMarker_M]a \\
            & \text{is feasible.}
    \end{split}
\end{align}
\end{subequations}
}

The users provided the following prompts:
\begin{itemize}
    \item $k=0$ User 1 says ``I want to have more CPUs!'' and User 3 says ``Please reduce the latency, please''.
    \item $k=1$ User 1 says ``Get more CPUs, please.''
    \item $k=2$ User 1 says ``I no longer need much CPU. Sorry!'', User 2 says ``May I use more CPUs, please? Thank you.'', and User 3 says ``Could I use more CPUs, please''.
\end{itemize}

\begin{figure*}[t]
  \begin{minipage}[b]{0.5\linewidth}
    \centering
    \subfigure[CPU usage at time $k=0$]{
        \includegraphics[width=\columnwidth]{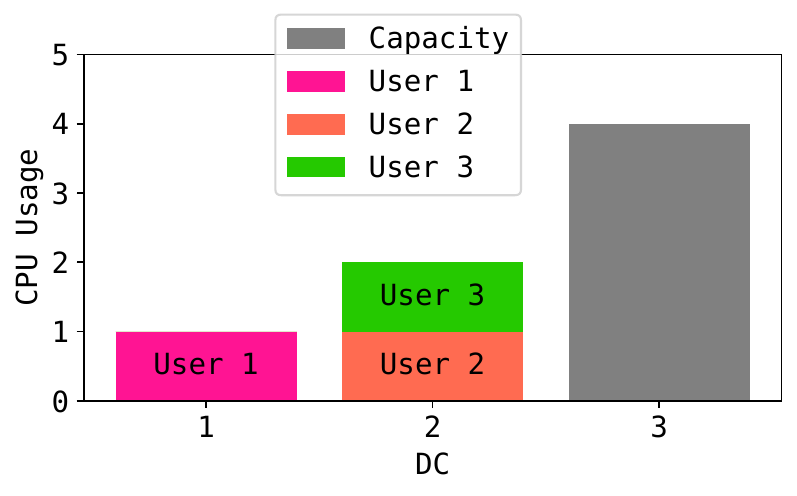}
        \label{F:E.D.CPUUsage.ChatStep0}
    }
  \end{minipage}
  \begin{minipage}[b]{0.5\linewidth}
    \centering
    \subfigure[CPU usage at time $k=1$]{
        \includegraphics[width=\columnwidth]{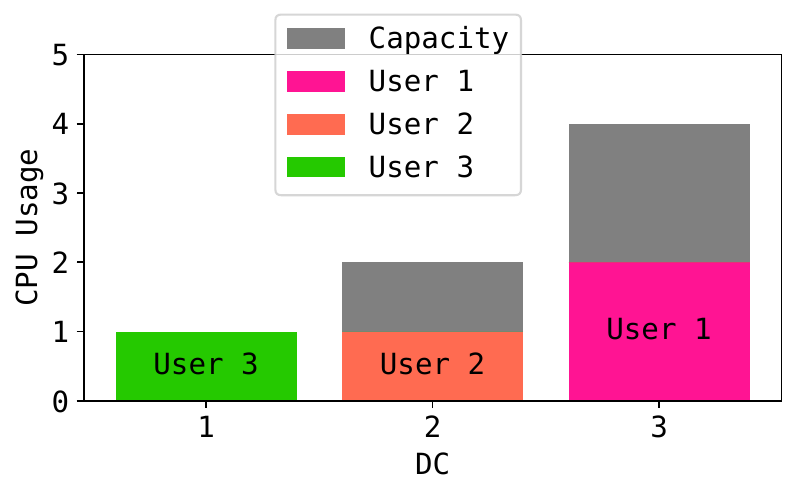}
        \label{F:E.D.CPUUsage.ChatStep1}
    }
  \end{minipage}
  \begin{minipage}[b]{0.5\linewidth}
    \centering
    \subfigure[CPU usage at time $k=2$]{
        \includegraphics[width=\columnwidth]{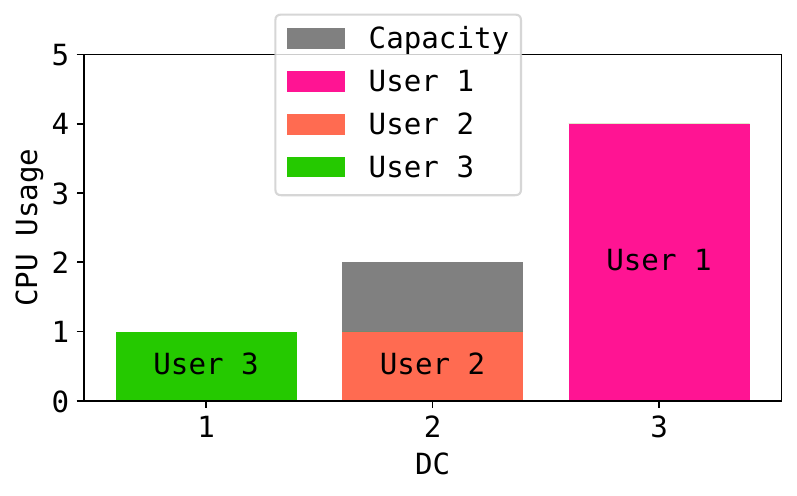}
        \label{F:E.D.CPUUsage.ChatStep2}
    }
  \end{minipage}
  \begin{minipage}[b]{0.5\linewidth}
    \centering
    \subfigure[CPU usage at time $k=3$]{
        \includegraphics[width=\columnwidth]{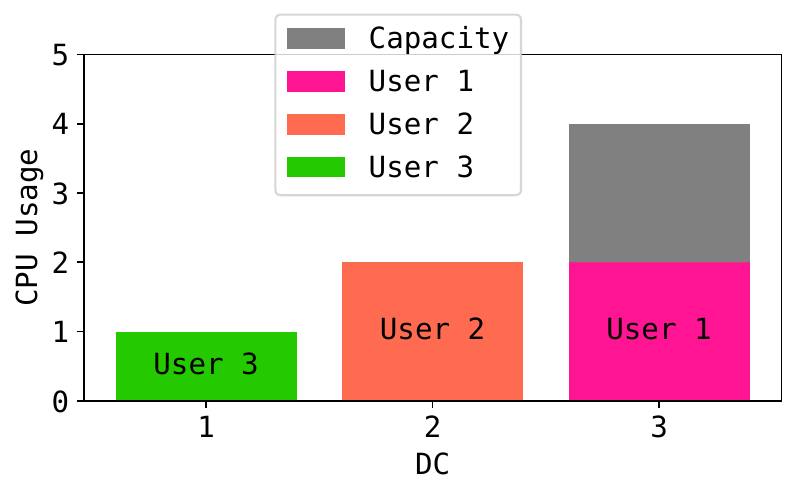}
        \label{F:E.D.CPUUsage.ChatStep3}
    }
  \end{minipage}
  
  \caption{CPU usage in the multi-user case}
  \label{F:E.D.Results}
\end{figure*}

The resulting solution $x$, which represents the VM assignment, is illustrated in Fig.~\ref{F:E.D.Results}. This figure displays the CPU usage of each DC. The figure illustrates how the system dynamically reallocates VMs to accommodate the multiple user requirements.

At chat step $k=0$, both User 1 and User 3's requirements are accepted. User 1's VM is moved to a DC with a larger available CPU capacity, and User 3's VM is moved from DC 2 to DC 1 for less latency.
User 1's requirement at chat step $k=1$ is also accepted.
\ColorB{At chat step $k=2$, the \textit{arbitration} is performed: User 1 and User 2 are accepted, and User 3 is rejected. }

\ColorB{By introducing an arbitration mechanism for simultaneously submitted user prompts, the network integrity is preserved.
Various methods for forming an arbitration exist; in this experiment, we search for the combination that accepts the largest number of users, but a scheme to distribute resources among all users can also be valid.}

\subsection{Intent Extractor Comparison}\label{Subsection:E.IntentExtractionComparison} 
This subsection evaluates the performance of the proposed intent extractor implementations. We compare the performance of different architectures: BERT+SVM and LLM. We also investigate the influence of the number of pre-labeled training samples on the intent extractor accuracy in each architecture.

We used Intel(R) Xeon(R) w5-2445 CPU and NVIDIA RTX A5000 GPU to perform the experiments.

A pre-labeled sample consists of a prompt and its expected update markers $[\CPUUpdateMarker~\LBUpdateMarker]$. A few samples are shown as follows:
\begin{itemize}
    \item ``Could you upgrade the CPU resource?'', $\CPUUpdateMarker=+1$, $\LBUpdateMarker=0$,
    \item ``My latency is too reduced'', $\CPUUpdateMarker=0$, $\LBUpdateMarker=-1$,
    \item ``I want to clean the system.'', $\CPUUpdateMarker=0$, $\LBUpdateMarker=0$.
\end{itemize}
Note that there are irrelevant prompts, and their expected update markers are $[\CPUUpdateMarker~\LBUpdateMarker]=[0~0]$. The other samples are shown in Appendix~\ref{Appendix:PreLabeledTrainingSamples}.

We evaluate the intent-extraction performance from the following perspectives:

\begin{description}[leftmargin=1em]
    \item[Change/No-Change Sensitivity] We treat  $\CPUUpdateMarker=0$ and $\LBUpdateMarker=0$ as negative labels and $\CPUUpdateMarker=-1,+1$ and $\LBUpdateMarker=-1,+1$ as positive labels, and measure the precision and recall metrics. This evaluation reflects how sensitively the interpreter reacts to prompts that imply parameter changes. 
    As shown in Table~\ref{T:E.D}, the results indicate that both BERT+SVM and LLM tend to have high recall but low precision. This suggests that the Interpreter is successful at capturing prompts that actually require a parameter change, but tends to over-interpret unrelated prompts as if they imply parameter changes.
    
    \item[Accuracy in Change Detection] We measure the balanced accuracy for $\CPUUpdateMarker=-1$ and $\CPUUpdateMarker=+1$ in the CPU topic, and $\LBUpdateMarker=-1$ and $\LBUpdateMarker=+1$ in the latency bound topic. This evaluation focuses on how accurately the Interpreter distinguishes between different directions of parameter changes.
    As shown in Table~\ref{T:E.D}, the LLM demonstrates higher classification performance than BERT+SVM with a small sample size (10 samples for the LLM), which suggests the LLM's adaptability to the domain.
    We also observe that classification accuracy can be extremely low under conditions with a small sample size (less than 10 for BERT+SVM and less than 5 for LLM).
\end{description}

As also shown in Table~\ref{T:E.D}, the LLM exhibited syntax errors, such as failing to produce an accurate JSON format in its response.
Furthermore, we can see that BERT+SVM requires an extremely shorter inference time compared to LLM, due to its smaller model size.

The following illustrates specific examples of the LLM intent extractor's response.

\begin{example}[Correct Intent Extraction]
The prompt is ``I want to have more CPUs !''. The response is as follows:

\textit{Here's the step-by-step thinking:\LineBreak\LineBreak The prompt is "I want to have more CPUs!". This sentence is straightforward and clearly indicates that the user wants to increase the number of CPUs.\LineBreak\LineBreak Now, let's analyze the intent:\LineBreak\LineBreak1. **Whether we should increase, decrease, or do nothing to allocate CPU resources.** The prompt explicitly states that the user wants to have more CPUs, so the intent is to **increase** the CPU resources.\LineBreak2. **Whether we should ease (increase), decrease, or do nothing to set the latency bound.** The prompt does not mention anything about latency, so we cannot determine the intent regarding the latency bound. Therefore, we set it to **none**.\LineBreak\LineBreak Here is the JSON code block that represents the user intent:\LineBreak ```\LineBreak \{\LineBreak "cpu": "increase",\LineBreak "latencybound": "none"\LineBreak \}\LineBreak```}

The response is parsed to $\UpdateMarker=[+1~0]$, which is correct.
\end{example}

\begin{example}[Incorrect Intent Extraction]
The prompt is ``I want to shutdown the virtual machine. How shold I do?''. The response is as follows:

\textit{**Step-by-step thinking:**\LineBreak\LineBreak The prompt is about shutting down a virtual machine, which is not directly related to CPU resources or latency bounds. However, we can infer some information from the context.\LineBreak\LineBreak The user is asking about how to shut down the virtual machine, which implies that they might be experiencing some issues or problems with the virtual machine. This could be related to performance or resource utilization.\LineBreak\LineBreak **Raw-text interpretation:**\LineBreak\LineBreak Based on the prompt, I'm going to assume that the user might be experiencing some performance issues with the virtual machine, which could be related to CPU resources. Since the user is asking about shutting down the virtual machine, it's likely that they want to free up resources.\LineBreak\LineBreak**JSON response:**\LineBreak```\LineBreak \{\LineBreak    "cpu": "decrease",\LineBreak    "latencybound": "none"\LineBreak \}\LineBreak```\LineBreak In this case, I'm assuming that the user wants to decrease CPU resources, as they're shutting down the virtual machine. The latency bound is set to "none" since the prompt doesn't provide any information about latency or network performance.}

The response is parsed to $\UpdateMarker=[-1 ~ 0]$, which is incorrect. The expected result is $\UpdateMarker=[0~0]$.
\end{example}

\begin{table*}
    \centering
    \caption{Intent Extractor Performance of CPU/Latency-bound topic}
    \begin{tabular}{ll|llllll}
    \hline
     Architecture & Training sample size & Computation time $\mathrm{[s]}$ & Precision & Recall & Balanced Accuracy & Syntax error $\mathrm{[\%]}$ \\
    \hline
    BERT+SVM & 30 & $5.40\times 10^{-3}$/$4.40\times 10^{-3}$ & 0.71/0.86 & 1.0/1.0 & 0.7/0.7 & 0/0 \\
    & 20 & $8.81\times 10^{-3}$/$5.12\times 10^{-3}$ & 0.67/0.86 & 1.0/1.0 & 0.93/0.7 & 0/0 \\
    & 10 & $5.10\times 10^{-3}$/$3.91\times 10^{-3}$ & 0.59/0.62 & 1.0/0.83 & 0/0.75 & 0/0 \\
    & 5 & $5.63\times 10^{-3}$/$4.64\times 10^{-3}$ & 0.59/0.5 & 1.0/1.0 & 0/0.9 & 0/0 \\
    & 3 & $4.92\times 10^{-3}$/$4.31\times 10^{-3}$ & 0.5/0.6 & 1.0/1.0 & 0.31/0 & 0/0 \\
    \hline
    LLM & 30 & $8.18$ & 0.91/0.6 & 1.0/1.0 & 1/1 & 0/0 \\
    & 20 & $7.7$ & 0.77/0.67 & 1.0/1.0 & 1/0.75 & 0/0 \\
    & 10 & $8.48$ & 0.59/0.6 & 1.0/1.0 & 1/0.75 & 7.1/7.1 \\
    & 5 & $8.38$ & 0.67/0.42 & 1.0/0.83 & 0.9/0 & 3.6/3.6 \\
    & 3 & $8.2$ & 0.62/0.42 & 1.0/0.83 & 1/0.75 & 3.6/3.6 \\
    & 0 (zero-shot) & $8.55$ & 0.67/0.4 & 1.0/1.0 & 0.9/0 & 3.6/3.6 \\
    \hline
    \end{tabular}
    
    \label{T:E.D}
\end{table*}

\section{Conclusion}\label{Section:C}
\ColorA{
We proposed a chat-driven network management framework that integrates an NLP-based interpreter with a conventional optimizer to enable chat-driven casual network reconfiguration.
This framework offers an iterative, loop process where user feedback in natural language updates the virtual network parameters.

The experimental results show that the proposed system effectively interprets \textit{linguistic} requests into \textit{structural} network adjustments, reallocating VMs and routing paths in both single and multi-user scenarios.
Performance evaluation of the intent extractor of the Interpreter is also performed. In this evaluation, the LLM exhibited superior accuracy in distinguishing change direction and adapting to the domain, while the lighter BERT+SVM model maintained a significantly shorter inference time, an important factor for real-time network operation.

The future works include the development of a conflict resolution and negotiation mechanism. Our multi-user scenario experiment demonstrated the system's ability to operate at the feasibility boundary, where conflicting requests led to infeasibility. 
Therefore, the next step is to design a communication mechanism to ensure that limited resources are shared fairly and to the user's satisfaction.
}

\smallskip
\noindent\textbf{Disclosure statement}: 
No potential conflict or interest was reported by the author(s).

\bibliographystyle{ieeetr}
\bibliography{Reference}

@inproceedings {Jacobs21_Lumi,author = {Arthur S. Jacobs and Ricardo J. Pfitscher and Rafael H. Ribeiro and Ronaldo A. Ferreira and Lisandro Z. Granville and Walter Willinger and Sanjay G. Rao},title = {{Hey, Lumi! Using Natural Language for Intent-Based Network Management}},booktitle = {2021 USENIX Annual Technical Conference (USENIX ATC 21)},year = {2021},isbn = {978-1-939133-23-6},pages = {625--639},url = {https://www.usenix.org/conference/atc21/presentation/jacobs},publisher = {USENIX Association},month = jul}

@article{Whaid14_IaaS,
title = {A survey on vehicular cloud computing},journal = {Journal of Network and Computer Applications},volume = {40},pages = {325-344},year = {2014},issn = {1084-8045},doi = {https://doi.org/10.1016/j.jnca.2013.08.004},url = {https://www.sciencedirect.com/science/article/pii/S1084804513001793},author = {Md Whaiduzzaman and Mehdi Sookhak and Abdullah Gani and Rajkumar Buyya},}

@inproceedings{Reimers19_SentenceBERT,title = "{Sentence-BERT: Sentence Embeddings using Siamese BERT-Networks}",  author = "Reimers, Nils and Gurevych, Iryna",  booktitle = "Proceedings of the 2019 Conference on Empirical Methods in Natural Language Processing",  month = "11",  year = "2019",  publisher = "Association for Computational Linguistics",  url = "https://arxiv.org/abs/1908.10084",}

@article{Dubey24_Llama3,title={{The Llama 3 Herd of Models}}, author={Abhimanyu Dubey and others},year={2024},eprint={2407.21783},archivePrefix={arXiv},primaryClass={cs.AI},url={https://arxiv.org/abs/2407.21783},  journal={arXiv preprint arXiv:2407.21783},}

@inproceedings{Summerhill06, author = {Rick Summerhill},title = {{The New Internet2 Network}},booktitle = {6th Annual Global LambdaGrid Workshop},year = {2006},}

@inproceedings{Abbas24_Leveraging,
  title={{Leveraging Large Language Models for Wireless Symbol Detection via In-Context Learning}},
  author={Abbas, Momin and Kar, Koushik and Chen, Tianyi},
  booktitle={Proceedings of 2024 IEEE Global Communications Conference (GLOBECOM)},
  pages={5217--5222},
  year={2024},
}

@inproceedings{Yao24_Velo,
  title={Velo: A vector database-assisted cloud-edge collaborative {LLM} Qo{S} optimization framework},
  author={Yao, Zhi and Tang, Zhiqing and Lou, Jiong and Shen, Ping and Jia, Weijia},
  booktitle={Proceedings of 2024 IEEE International Conference on Web Services (ICWS)},
  pages={865--876},
  year={2024},
}

@inproceedings{Su24_Leveraging,
  title={{Leveraging Large Language Models for VNF Resource Forecasting}},
  author={Su, Jing and Nair, Suku and Popokh, Leo},
  booktitle={Proceedings of 2024 IEEE 10th International Conference on Network Softwarization (NetSoft)},
  pages={258--262},
  year={2024},
}

@inproceedings{Manias24_Semantic,
  title={{Semantic Routing for Enhanced Performance of LLM-Assisted Intent-Based 5G Core Network Management and Orchestration}},
  author={Manias, Dimitrios Michael and Chouman, Ali and Shami, Abdallah},
  booktitle={GLOBECOM 2024-2024 IEEE Global Communications Conference},
  pages={2924--2929},
  year={2024},
  organization={IEEE}
}

@article{Mcnamara23_NLP,
  title={{NLP} powered intent based network management for private {5G} networks},
  author={Mcnamara, Joseph and Camps-Mur, Daniel and Goodarzi, Meysam and Frank, Hilary and Chinchilla-Romero, Lorena and Ca{\~n}ellas, Ferr{\'a}n and Fern{\'a}ndez-Fern{\'a}ndez, Adriana and Yan, Shuangyi},
  journal={IEEE Access},
  volume={11},
  pages={36642--36657},
  year={2023},
}

@article{Asif25_RIBN,
  title={{R-IBN:} A reinforcement learning-based intent-driven framework for end-to-end service orchestration and optimization},
  author={Asif, Muhammad and Khan, Talha Ahmed and Song, Wang-Cheol},
  journal={Computer Networks},
  pages={111564},
  year={2025},
  publisher={Elsevier}
}

@inproceedings{Habib25_LLM,
  title={{LLM}-based intent processing and network optimization using attention-based hierarchical reinforcement learning},
  author={Habib, Md Arafat and Rivera, Pedro Enrique Iturria and Ozcan, Yigit and Elsayed, Medhat and Bavand, Majid and Gaigalas, Raimundus and Erol-Kantarci, Melike},
  booktitle={Proceedings of 2025 IEEE Wireless Communications and Networking Conference (WCNC)},
  pages={1--6},
  year={2025},
}

@inproceedings{Lotfi25_LLM,
  title={{LLM}-augmented deep reinforcement learning for dynamic {O-RAN} network slicing},
  author={Lotfi, Fatemeh and Rajoli, Hossein and Afghah, Fatemeh},
  booktitle={Proceedings of 2025 IEEE International Conference on Communications (ICC)},
  year={2025}
}

@article{Yi18_Comprehensive,
  title={A comprehensive survey of network function virtualization},
  author={Yi, Bo and Wang, Xingwei and Li, Keqin and Huang, Min and others},
  journal={Computer Networks},
  volume={133},
  pages={212--262},
  year={2018},
  publisher={Elsevier}
}

@article{Schardong21_NFV,
  title={{NFV} resource allocation: A systematic review and taxonomy of VNF forwarding graph embedding},
  author={Schardong, Frederico and Nunes, Ingrid and Schaeffer-Filho, Alberto},
  journal={Computer Networks},
  volume={185},
  pages={107726},
  year={2021},
  publisher={Elsevier}
}

@inproceedings{Huin17_Optimization,
  title={Optimization of network service chain provisioning},
  author={Huin, Nicolas and Jaumard, Brigitte and Giroire, Fr{\'e}d{\'e}ric},
  booktitle={Proceedings of 2017 IEEE International Conference on Communications (ICC)},
  pages={1--7},
  year={2017},
}

@inproceedings{Urata24_Distributionally,
  title={Distributionally Robust Virtual Network Allocation Under Uncertainty of Renewable Energy Power with Wasserstein Metric},
  author={Urata, Kengo and Nakamura, Ryota and Harada, Shigeaki},
  booktitle={Proceedings of 2024 IEEE International Conference on Communications (ICC2024)},
  pages={2859--2864},
  year={2024}
}

@inproceedings{Alameddine17_Scheduling,
  title={Scheduling service function chains for ultra-low latency network services},
  author={Alameddine, Hyame Assem and Qu, Long and Assi, Chadi},
  booktitle={Proceedings of 13th international conference on network and service management (CNSM)},
  pages={1--9},
  year={2017},
}

@inproceedings{Carpio17_VNF,
  title={VNF placement with replication for Loac balancing in NFV networks},
  author={Carpio, Francisco and Dhahri, Samia and Jukan, Admela},
  booktitle={Proceedings of 2017 IEEE international conference on communications (ICC)},
  pages={1--6},
  year={2017},
}

@article{Hosseini19_Probabilistic,
  title={Probabilistic virtual link embedding under demand uncertainty},
  author={Hosseini, Fatemeh and James, Alexander and Ghaderi, Majid},
  journal={IEEE Transactions on Network and Service Management},
  volume={16},
  number={4},
  pages={1552--1566},
  year={2019},
}

@inproceedings{Van24_Towards,
  title={Towards intent-based configuration for network function virtualization using in-context learning in large language models},
  author={Van Tu, Nguyen and Yoo, Jae-Hyoung and Hong, James Won-Ki},
  booktitle={Proceedings of 2024 IEEE Network Operations and Management Symposium (NOMS)},
  pages={1--8},
  year={2024},
}

@article{Tu25_Intent,
  title={Intent-Based Network Configuration Using Large Language Models},
  author={Tu, Nguyen and Nam, Sukhyun and Hong, James Won-Ki},
  journal={International Journal of Network Management},
  volume={35},
  number={1},
  pages={e2313},
  year={2025},
  publisher={Wiley Online Library}
}

@inproceedings{Houssel24_Towards,
  title={Towards explainable network intrusion detection using large language models},
  author={Houssel, Paul RB and Singh, Priyanka and Layeghy, Siamak and Portmann, Marius},
  booktitle={Proceedings of 2024 IEEE/ACM International Conference on Big Data Computing, Applications and Technologies (BDCAT)},
  pages={67--72},
  year={2024},
}

@article{Rezaei25_Fedllmguard,
  title={{FedLLMGuard}: A federated large language model for anomaly detection in 5G networks},
  author={Rezaei, Hadiseh and Taheri, Rahim and Shojafar, Mohammad},
  journal={Computer Networks},
  pages={111473},
  year={2025},
  publisher={Elsevier}
}

@article{Daniels09_Server,
  title={Server virtualization architecture and implementation},
  author={Daniels, Jeff},
  journal={XRDS: Crossroads, The ACM Magazine for Students},
  volume={16},
  number={1},
  pages={8--12},
  year={2009},
  publisher={ACM New York, NY, USA}
}

@inproceedings{Miyaoka24_Chatmpc,
  title={{ChatMPC}: Natural language based mpc personalization},
  author={Miyaoka, Yuya and Inoue, Masaki and Nii, Tomotaka},
  booktitle={Proceedings of 2024 IEEE American Control Conference (ACC)},
  pages={3598--3603},
  year={2024},
  organization={IEEE}
}

@inproceedings{Miyaoka25_ChatDriven,
  title={Chat-Driven Interface for Virtual Network Reallocation},
  author={Miyaoka, Yuya and Inoue, Masaki and Urata, Kengo and Harada, Shigeaki},
  booktitle={Proceedings of 2025 IEEE International Conference on Communications (ICC2025)},
  pages={1608--1613},
  year={2025},
  organization={IEEE}
}

\appendix

\subsection{Pre-labeled Training Samples}\label{Appendix:PreLabeledTrainingSamples}
Table~\ref{T:Appendix.PLTS.PLTS} shows the prompts and their expected update markers $[\CPUUpdateMarker~\LBUpdateMarker]$ used as pre-labeled training samples in Section~\ref{Section:E}.

\begin{table}[t]
    \centering
    \caption{Pre-labeled training samples}
    \begin{tabular}{lll}
\hline
Prompt & $\CPUUpdateMarker$ & $\LBUpdateMarker$ \\
\hline
Could you upgrade the CPU resource? & $+1$ & $0$ \\
Upgrade my CPU resource, please & $+1$ & $0$ \\
I want more CPU & $+1$ & $0$ \\
I need more CPUs & $+1$ & $0$ \\
The current CPU is not sufficient & $+1$ & $0$ \\
My CPU is underperforming & $+1$ & $0$ \\
The user needs a CPU upgrade & $+1$ & $0$ \\
Could you reduce the number of CPUs? & $-1$ & $0$ \\
I'm requesting a decrease in CPU resource & $-1$ & $0$ \\
It is ok to decrease the CPU. & $-1$ & $0$ \\
I want less CPU. & $-1$ & $0$ \\
The number of CPUs is more than I can afford. & $-1$ & $0$ \\
Could you offer fewer CPUs? & $-1$ & $0$ \\
The user need to free up some CPU capacity. & $-1$ & $0$ \\
\hline
Could you reduce the latency of my network? & $0$ & $-1$ \\
Reduce the latency of my network, please & $0$ & $-1$ \\
I need less latency & $0$ & $-1$ \\
I need lower latency network & $0$ & $-1$ \\
My network speed is so late & $0$ & $-1$ \\
My network is too late & $0$ & $-1$ \\
The user's latency is late & $0$ & $-1$ \\
I don't want such a low latency network & $0$ & $+1$ \\
I don't need such a low latency network & $0$ & $+1$ \\
My latency is too reduced & $0$ & $+1$ \\
The user do not need less latency & $0$ & $+1$ \\
The user's latency is too reduced & $0$ & $+1$ \\
It is ok to ease the latency bound & $0$ & $+1$ \\
\hline
Could you tell me what time the party starts? & $0$ & $0$ \\
I want to clean the system. & $0$ & $0$ \\
My cat prefers walking rather than running. & $0$ & $0$ \\
No one would comes. & $0$ & $0$ \\
May I come in? & $0$ & $0$ \\
How do I know the sudo password? & $0$ & $0$ \\
\hline
\end{tabular}
    \label{T:Appendix.PLTS.PLTS}
\end{table}

\subsection{LLM Input Template}\label{Appendix:LLMInputTemplate}
\begin{lstlisting}[breaklines]
### Instruction

You are an operator of a network infrastructure service.
You need to extract the user's intent from the prompt they provide. The intent data is used to adapt the network settings to the user's preference.
Please interpret the prompt from two aspects:

1. **Whether we should increase, decrease, or do nothing to allocate CPU resources.**
2. **Whether we should ease (increase), decrease, or do nothing to set the latency bound.**
The latency bound is the constraint that sets the maximum latency time of the network communication.
The lower the latency bound, the faster the user experiences the network.
The more latency bound is eased (relaxed or mitigated), the reasonable and flexible (with more room to enforce other requirements more strictly) the user can afford to be.

Please note that
* The prompt may be uninterpretable: it may not include any CPU-related topics or latency-bound topics.
* The prompt may not give a direct statement of user intent, and we may have to interpret the intent step-by-step.
* Your response needs to include a JSON code block that represents the user intent.

### Response Format
You can include the step-by-step thinking in raw-text format.
After that, you must include the following JSON format:
```
{
    "cpu":"increase",
    "latencybound":"ease"
}
```
* The `cpu` value must be set to "increase", "decrease", or "none". Choose "none" if the prompt is uninterpretable.
* The `latencybound` value must be set to "reduce", "ease", or "none". Choose "none" if the prompt is uninterpretable.
{% if example %}
### Example
{{example}}
{% endif %}
### Prompt
{{prompt}}
\end{lstlisting}

\end{document}